\documentclass{aip-cp}

\usepackage[numbers]{natbib}
\usepackage{rotating}
\usepackage{graphicx}

\begin{document}

\title{Quantification of the electric dipole moment generated by hadronic CP violation: resolution of the strong CP problem}

\author[aff1,aff2]{Nodoka Yamanaka\corref{cor1}}
\affil[aff1]{
Kobayashi-Maskawa Institute for the Origin of Particles and the Universe, Nagoya University, Furocho, Chikusa, Aichi 464-8602, Japan
}
\affil[aff2]{
Nishina Center for Accelerator-Based Science, RIKEN, Wako 351-0198, Japan
}
\corresp[cor1]{Corresponding author: nyamanaka@kmi.nagoya-u.ac.jp}

\maketitle

\begin{abstract}
The atomic, nuclear, and nucleon electric dipole moments (EDMs) have significant sensitivity to the CP violation of elementary particle physics, but its quantification has for long been obstructed by the nonperturbative physics of quantum chromodynamics.
Quite recently, there were significant progresses in this field, notably the quantification of hadron level CP violation and the resolution of the strong CP problem.
In this proceedings contribution, we summarize this attempt by focussing on the latter topic, the resulting new paradigm of baryogenesis, and the modifications that have to be applied in the evaluation of the EDMs.
\end{abstract}

\section{INTRODUCTION}

As stated by Sakharov, the CP violation is required to realize the matter abundant Universe \cite{Sakharov:1967dj}.
The atomic, nuclear, and nucleon electric dipole moments (EDMs) have significant sensitivity to the CP violation of elementary particle physics \cite{Pospelov:2005pr,Yamanaka:2014mda,Yamanaka:2016umw,Yamanaka:2017mef,Chupp:2017rkp,Alarcon:2023gfu}, and they are the most attractive experimental observables thanks to the very high accuracy \cite{Graner:2016ses,Abel:2020gbr,Roussy:2022cmp} and the very low standard model (SM) background \cite{Seng:2014lea,Yamanaka:2015ncb,Yamanaka:2016fjj,Lee:2018flm,Yamaguchi:2020eub,Yamaguchi:2020dsy,Ema:2022yra}.

Unless the EDM of an elementary particle is directly measured, the nonperturbative physics of quantum chromodynamics (QCD) always becomes problematic in extracting the CP violation of the fundamental theory, since the hadronic intermediate states always contribute to composite systems.
Recently, there were significant progresses in this field, notably the resolution of the strong CP problem and the quantification of hadron level CP violation.

For determining hadronic CP-odd effective interactions, we first have to calculate the renormalization group evolution because CP-odd quark-gluon operators mix with each other due to QCD corrections \cite{Degrassi:2005zd,Hisano:2012cc}.
After obtaining the Wilson coefficients at the hadronic scale, we then have to match the QCD operators with hadron level interactions.
The ideal matching is to calculate hadronic matrix elements in lattice QCD.
However, in the context of the EDM, the only quantified one is the nucleon tensor charge which corresponds to the quark EDM contribution to the nucleon EDM \cite{Yamanaka:2018uud,FlavourLatticeAveragingGroupFLAG:2021npn}.
The quark chromo-EDM \cite{Pospelov:2001ys,deVries:2016jox,Sahoo:2023vrn} and the CP-odd four-quark contributions \cite{An:2009zh} can be evaluated using chiral perturbation theory in combination with QCD sum rules and the large $N_c$ factorization, respectively.
The effect of the Weinberg operator (CP violating three-gluon interaction) which was so far difficult to access, could also recently be estimated \cite{Yamanaka:2020kjo,Osamura:2022rak,Yamanaka:2022qlu}, although it still involves a large uncertainty.

At a more fundamental level, the hadronic CP violation is affected by an important problem due to the topological configurations of gauge fields.
Formally, QCD allows a CP violating interaction term (the so-called $\theta$-term),
\begin{equation}
{\cal L}_{\theta}
=
\theta N_f \frac{\alpha_s}{8\pi} F_{\mu \nu, a} \tilde F^{\mu \nu}_a
,
\label{eq:theta-term}
\end{equation}
where $F^{\mu \nu}_a$ is the field strength of the gauge field $A^\mu_a(x)$ and $\tilde F^{\mu \nu}_a$ its dual, but it is known that the experimental data of the EDMs of neutron and atoms are strongly constraining it, as $|\theta | < 10^{-10}$ \cite{Abel:2020gbr,Crewther:1979pi,Graner:2016ses,Yanase:2020agg,Yanase:2020oos,Yanase:2023jdr,Hubert:2022pnl}.
This extreme fine-tuning is nowadays called the {\it Strong CP problem}. 
Quite recently, this issue was claimed to be resolved within QCD by several groups \cite{Ai:2020ptm,Nakamura:2021meh,Yamanaka:2022vdt,Yamanaka:2022bfj}.
The statement of Refs. \cite{Yamanaka:2022vdt,Yamanaka:2022bfj}, that the topological charge is unphysical, is particularly strong, and has a lot of important consequences in particle physics, notably in the contexts of the baryogenesis and the EDM.
In this proceedings contribution, we review the resolution of the strong CP problem and its consequences.
We first briefly show the mechanism of the unobservability of topological charge, then the constraints on baryogenesis that result from it, and finally the modifications that have to be applied in the EDM calculations.
The last section is devoted to the conclusion.

\section{UNPHYSICAL TOPOLOGICAL CHARGE}

Nonabelian gauge fields are known to have nontrivial configurations that can be classified using integer numbers.
The topological charge of the gauge field is defined as follows:
\begin{equation}
\int d^4 x \,
\frac{\alpha_s}{8\pi} F_{\mu \nu, a} \tilde F^{\mu \nu}_a
=
\frac{i g_s \alpha_s}{24\pi}
\int d^3 x \,
f_{abc}
\epsilon_{ijk}
A_{i,a}(x)A_{j,b}(x)A_{k,c}(x)
\Bigg|^{t=+\infty}_{t=-\infty}
=
\Delta n
.
\label{eq:topological_charge}
\end{equation}
The integrand is actually a total divergence, but its total volume integral may be a non-zero integer number $\Delta n$, which is due to the topologically nontrivial ``winding'' of the gauge field at infinite time.
The essential point of the unobservability of this quantity is the contraction of three gauge field operators with the Levi-Civita tensor $\epsilon_{ijk}$ which covers all three-dimensional directions.
Since the time dependence is frozen at infinite time, the gauge fields have only three degrees of freedom, and the contraction with $\epsilon_{ijk}$ necessarily involves the unphysical longitudinal gauge component.
The unobservability of the longitudinal gauge field is certified by the BRST symmetry, which always holds in perturbation theory.
The perturbativity is warranted by the Adler-Bardeen theorem \cite{Adler:1969er} which forbids any higher-order corrections to the topological charge density $\frac{\alpha_s}{8\pi} F_{\mu \nu, a} \tilde F^{\mu \nu}_a$.
This can be understood by the fact that the topological charge is an integer number, which cannot be shifted by continuous radiative corrections.
The unobservability of the topological charge can also be shown in another approach using the Ward-Takahashi identity (WTI) of BRST symmetry \cite{Yamanaka:2022bfj}.
Since the $\theta$-term (\ref{eq:theta-term}) is also defined by the topological charge, its unphysicalness immediately implies that the strong CP problem is resolved.

\section{BARYOGENESIS REVISITED\label{sec:baryogenesis}}

We now include the fermion.
The chiral WTI reads \cite{Bardeen:1969md}
\begin{equation}
\sum_{\psi}^{N_f}
\Bigl[
\partial^\mu (\bar \psi \gamma_\mu \gamma_5 \psi )
+2m_\psi
\bar \psi i\gamma_5 \psi
\Bigr]
=
-
N_f
\frac{\alpha_s}{8\pi} F_{\mu \nu, a}\tilde F^{\mu \nu}_a
,
\label{eq:chiralWTI}
\end{equation}
where the left-hand side contains the axial current of the fermion $\psi$ and the mass contribution, while the nonzero right-hand side means that the conservation of the axial current is violated by the topological charge density.
However, we saw in the previous section that the topological charge is unphysical.
So what is the unphysical part of the fermion?
The topological charge is actually related to the chiral Dirac zero-modes (eigenfunctions of the Dirac operator $D\hspace{-0.55em}/\,$ with zero eigenvalue) via the Atiyah-Singer theorem:
\begin{equation}
{\rm ind}(D\hspace{-0.55em}/\,)
=
-N_f 
\int d^4 x\,
\frac{\alpha_s}{8\pi} F_{\mu \nu, a}\tilde F^{\mu \nu}_a
.
\label{eq:atiyah-singer}
\end{equation}
The unphysical part of the fermion is then the chiral Dirac zero-modes.
Removing the unphysical components from both sides of Eq. (\ref{eq:chiralWTI}), we obtain the following ``physical chiral WTI''
\begin{equation}
\sum_\psi^{N_f}
\Bigr[
\partial^\mu (\bar \psi \gamma_\mu \gamma_5 \psi )
+2m_\psi
\bar \psi i\gamma_5 \psi
\Bigr]_{\lambda \ne 0}
=
-N_f \frac{\alpha_s}{8\pi} F_{\mu \nu, a}\tilde F^{\mu \nu}_a
\Bigr|_{\Delta n =0}
,
\label{eq:physicalchiralWTI}
\end{equation}
where $\lambda \ne 0$ means that we removed the chiral Dirac zero-modes.
The physical chiral WTI simply means that the axial $U(1)$ current is not violated by the quantum anomaly and that it is conserved up to the fermion mass.

In the SM, it was known so far that the baryon and lepton numbers ($B+L$) are anomalously violated, and it was believed that the chiral Dirac zero-modes can combine with each other to form the baryon and lepton number violating 't Hooft vertex \cite{tHooft:1976rip}, typically generating processes like
\begin{equation}
u + d 
\to
\bar d + \bar s + 2\bar c +3\bar t + e^+ + \mu^+ + \tau^+
.
\end{equation}
The generation of baryon number through the high temperature sphaleron process in the early cosmological era was an attractive scenario to explain the current matter abundance of the Universe.

However, the above discussion forbids any 't Hooft vertices, since the chiral Dirac zero-modes are unphysical.
The SM is therefore baryon and lepton number preserving.
By also noting that recent experimental data suggest that the Higgs sector does not have a first order phase transition \cite{Kajantie:1996mn} and that the CP violation due to the Cabibbo-Kobayashi-Maskawa (CKM) matrix \cite{Kobayashi:1973fv} is suppressed by the Glashow-Iliopoulos-Maiani (GIM) mechanism \cite{Glashow:1970gm}, none of the three criteria of Sakharov \cite{Sakharov:1967dj} are satisfied by the SM and it is ``triply'' impossible to realize the matter abundant Universe within it.
At the same time, the presence of matter around us is a direct proof of the existence of explicit local baryon number violating interactions beyond the SM.
Good candidates to be investigated are the grand unified theory \cite{Georgi:1974sy}, leptoquark models \cite{Dorsner:2016wpm}, or R-parity violating supersymmetry \cite{Barbier:2004ez}.

\section{CP-ODD MASS OF FERMIONS AND EDM}

The chiral WTI (\ref{eq:chiralWTI}) implied so far that the $\theta$-term and the CP-odd mass of fermions
\begin{equation}
{\cal L}_{\rm odd}
=
-m_{\rm odd} \bar \psi i \gamma_5 \psi
,
\label{eq:CP-odd_mass}
\end{equation}
can be converted to each other.
However, we saw that the topological charge is unphysical, so the CP-odd fermion mass is now free to define and can then be rotated away.
This is an important consequence of the physical chiral WTI (\ref{eq:physicalchiralWTI}) which states that the anomalous breaking of chiral $U(1)$ symmetry is not observable.

If $\theta$ is physical, the CP-odd quark mass contributes to the EDMs of neutron and atoms, so there must have been strong constraints on candidates of new physics beyond the SM which generate Eq. (\ref{eq:CP-odd_mass}) through loops, such as the supersymmetric models \cite{Pospelov:2005pr}.
These limits are now nonexisting since both $\theta$ and the CP-odd quark mass are unphysical.
This also implies that the CP phase of the CKM matrix is the only source of CP violation in the SM.

It is interesting to note that the CP-odd quark mass and the $\theta$-term do not decouple even if the scale of new physics is raised, and they were problematic in the decoupling of high energy physics.
The above discussion resolves this problem since these dangerous non-decoupling terms are actually not observable.

Superficially, our resolution seems to exhibit similar effects as the well-known axion mechanism \cite{DiLuzio:2020wdo} which minimizes the $\theta$-term and the CP-odd fermion mass, but there are also differences.
The leading order difference is of course that we do not need to introduce additional fields, such as the axion.
As a minor difference, we have the absence of the ``induced'' $\theta$-term, which is generated by the shift of the vacuum by the following correlation
\begin{equation}
K_1 =
-i N_f \frac{\alpha_s}{8\pi}
\lim_{k \to 0} \int d^4 x\, e^{ik\cdot x} \langle 0 | T\{ F_{\mu \nu, a}\tilde F^{\mu \nu}_a (x) O_{CP}(0) \} | 0 \rangle
,
\label{eq:linearaxioncoupling}
\end{equation}
where $O_{CP}$ is some CP-odd quark-gluon operator.
This contribution vanishes because the operator $\frac{\alpha_s}{8\pi} F_{\mu \nu, a}\tilde F^{\mu \nu}_a$ becomes isolated in the zero-momentum limit and becomes the unphysical topological charge.

\section{CONCLUSION}

In this proceedings contribution, we presented the argument that the topological charge of QCD is not observable and that the strong CP problem is actually resolved.
The strong CP problem was for long a problem in the context of the hadronic CP violation.
Its resolution now allows us quantifiable analyses of new physics beyond the SM using EDMs.
We also saw that the sphaleron induced baryogenesis which was an attractive candidate scenario to explain the matter abundance of the Universe is now forbidden.
We therefore need to look for explicit local baryon number violating interactions in models such as the grand unified theory.

Now what is the next task left for the quantification of the EDM?
The hadronic matrix element which needs to be accurately determined is presumably the pion-nucleon sigma-term ($\sigma_{\pi N}$), which currently has a large discrepancy between results of lattice calculations \cite{Yamanaka:2018uud,FlavourLatticeAveragingGroupFLAG:2021npn,Gupta:2021ahb} and phenomenological extractions \cite{Hoferichter:2023ptl}.
It is not only an important input in the chiral perturbation theory of CP-odd pion-nucleon interaction, but it also yields the CP-odd electron-nucleon interaction \cite{Yanase:2018qqq} which is one of the leading contribution to the atomic EDM.


\section{ACKNOWLEDGMENTS}
This work was supported by Daiko Foundation.


\nocite{*}
\bibliographystyle{aipnum-cp}%
\bibliography{yamanaka}%

\end{document}